\documentclass[aps,prb,superscriptaddress,floatfix,twocolumn,amsmath,amssymb,showpacs]{revtex4-1}

\usepackage{graphicx}
\usepackage{bm}
\usepackage{color}
\usepackage{epstopdf}

\begin{document}
\title{The role of atomic vacancies and boundary conditions on ballistic thermal transport in graphene nanoribbons}
\author{P. Scuracchio}
\affiliation{
Facultad de Ciencias Exactas Ingenier{\'\i}a y Agrimensura, Universidad Nacional de Rosario and Instituto de
F\'{\i}sica Rosario, Bv. 27 de Febrero 210 bis, 2000 Rosario,
Argentina.}
\author{S. Costamagna}
\affiliation{
Facultad de Ciencias Exactas Ingenier{\'\i}a y Agrimensura, Universidad Nacional de Rosario and Instituto de
F\'{\i}sica Rosario, Bv. 27 de Febrero 210 bis, 2000 Rosario,
Argentina.}
\affiliation{Universiteit Antwerpen,
Groenenborgerlaan 171, BE-2020 Antwerpen, Belgium.
}
\author{F. M. Peeters}
\affiliation{Universiteit Antwerpen,
Groenenborgerlaan 171, BE-2020 Antwerpen, Belgium.
}
\author{A. Dobry}
\affiliation{
Facultad de Ciencias Exactas Ingenier{\'\i}a y Agrimensura, Universidad Nacional de Rosario and Instituto de
F\'{\i}sica Rosario, Bv. 27 de Febrero 210 bis, 2000 Rosario,
Argentina.}
\date{\today}


\begin{abstract}

Quantum thermal transport in armchair and zig-zag graphene nanoribbons are investigated in the presence of single atomic vacancies and subject to different boundary conditions. 
We start with a full comparison of the phonon polarizations and energy dispersions
as given by a fifth-nearest-neighbor force-constant model (5NNFCM)
and by elasticity theory of continuum membranes (ETCM). 
For free-edges ribbons we discuss the behavior of an additional acoustic edge-localized flexural mode, 
known as fourth acoustic branch (4ZA), which has a small gap when it is obtained by the 5NNFCM.
Then, we show that ribbons with supported-edges have a sample-size dependent energy gap in the phonon spectrum
which is particularly large for in-plane modes.
Irrespective to the calculation method and the boundary condition, the dependence of the energy gap for the low-energy 
optical phonon modes against the ribbon width W is found to be proportional to 1/W for in-plane, 
and 1/W$^2$ for out-of-plane phonon modes.
Using the 5NNFCM, the ballistic thermal conductance and its contributions from every single phonon mode are then
obtained by the non equilibrium Green's function technique. 
We found that, while edge and central localized single atomic vacancies do not affect the low-energy 
transmission function of in-plane phonon modes, they reduce considerably the contributions of the flexural modes.
On the other hand, in-plane modes contributions are strongly dependent on the boundary conditions 
and at low temperatures can be highly reduced in supported-edges samples.
These findings could open a route to engineer graphene based devices where it 
is possible to discriminate the relative contribution of polarized 
phonons and to tune the thermal transport on the nanoscale.       

\end{abstract}


\maketitle


\section{Introduction}

The first astonishing properties seen in graphene after its exfoliation were connected 
with its electronic structure. The anomalous quantum Hall effect and the ultra-high 
electron mobility\cite{Novo,Hall} are among the most celebrated properties. 
They are consequences of the electronic band structure characterized 
by two inequivalent cones at the edges of the Brilloiun zone
making electrons behave as ultra-relativistic massless Dirac fermions\cite{Novo}.
Later on, other fascinating properties were discovered. 
Graphene is now considered as a multi-functional material, combining outstanding electric, optical, mechanical and thermal properties\cite{Opt1,Opt2,Mec1,Mec2,Seol,Balandin1,Ghosh}.  
In particular, thermal transport plays an exciting role in graphene physics. 
Measurements have shown that graphene is one of the best 
heat conductors with a thermal conductivity $\kappa$ as high as $\sim$ 5000~W/mK 
in suspended samples\cite{Seol, Balandin1, Ghosh}. 
These results opened the route for new thermal 
control applications in nanoelectronics\cite{App1,App2,App3,App4,Pop,Trans}. 

Numerous theoretical studies have addressed the problem of the thermal conductivity in graphene 
but the intrinsic mechanisms behind the large value of $\kappa$ remain still to be 
clarified\cite{Balandin1, Balandin2, Balandin3,TranspRev,Xu}.
Within graphene based nanosctructures, nanoribbons (GNRs) are among the most studied. 
Their particular geometry and possible configurational defects, such as vacancies or $^{13}$C carbon isotopes, 
modify the usual phonon spectrum and therefore the thermal transport properties.\cite{Chen,Hu,Jiang,Zhang4,Haskings,Pop2,Chen}.
Understanding the microscopic mechanism for thermal transport in confined geometries is therefore crucial to control the 
heat dissipation at the nanoscale. Although some insights of GNRs thermal properties can be 
deduced from former studies on carbon nanotubes\cite{Wang,CNT}, to our knowledge, 
detailed microscopic explanation of the effect of boundary conditions and atomic vacancies in ribbons is still lacking.

In the present paper we study the vibrational characteristics and the 
thermal transport properties of graphene nanoribbons with single atomic vacancies and subjected 
to different boundary conditions: free- and supported-edges. 
To this end we adopt a fifth-nearest-neighbor force-constant model (5NNFCM)\cite{Michel,Mohr,Twisted}
to obtain the phonon spectrum of both, armchair (AGNR) and zig-zag (ZGNR) ribbons, and compare the results with those obtained by the elasticity theory of continuum membranes (ETCM) \cite{Landau,grapheneacoustic, Scuracchio}. Thermal transport properties are then addressed by means of the nonequilibrium Green's function technique (NEGF)\cite{Mingo,Huang}. We analyze here 
the contributions from in-plane and out-of-plane phonon modes to the ballistic thermal conductance. 
Our results show that it is possible to control the partial contributions of polarized phonons to the
thermal conductance of GNRs.  

The paper is organized as follows. In Section II we compare the out-of-plane phonon modes of GNRs as determined by the ETCM and the lattice dynamic 5NNFCM. In Section III, we use the 5NNFCM within the NEGF technique to calculate polarized phonon transmissions and thermal conductance of free- and supported-edges ZGNRs in the presence of single vacancies. Finally, in Section VI we discuss the conclusions and perspectives of our work.


\section{Phonon modes}

\begin{figure}[t]
\vspace{0.6cm}
\includegraphics[trim = 0cm 21cm 0cm 0.05cm, clip, width=18pc]{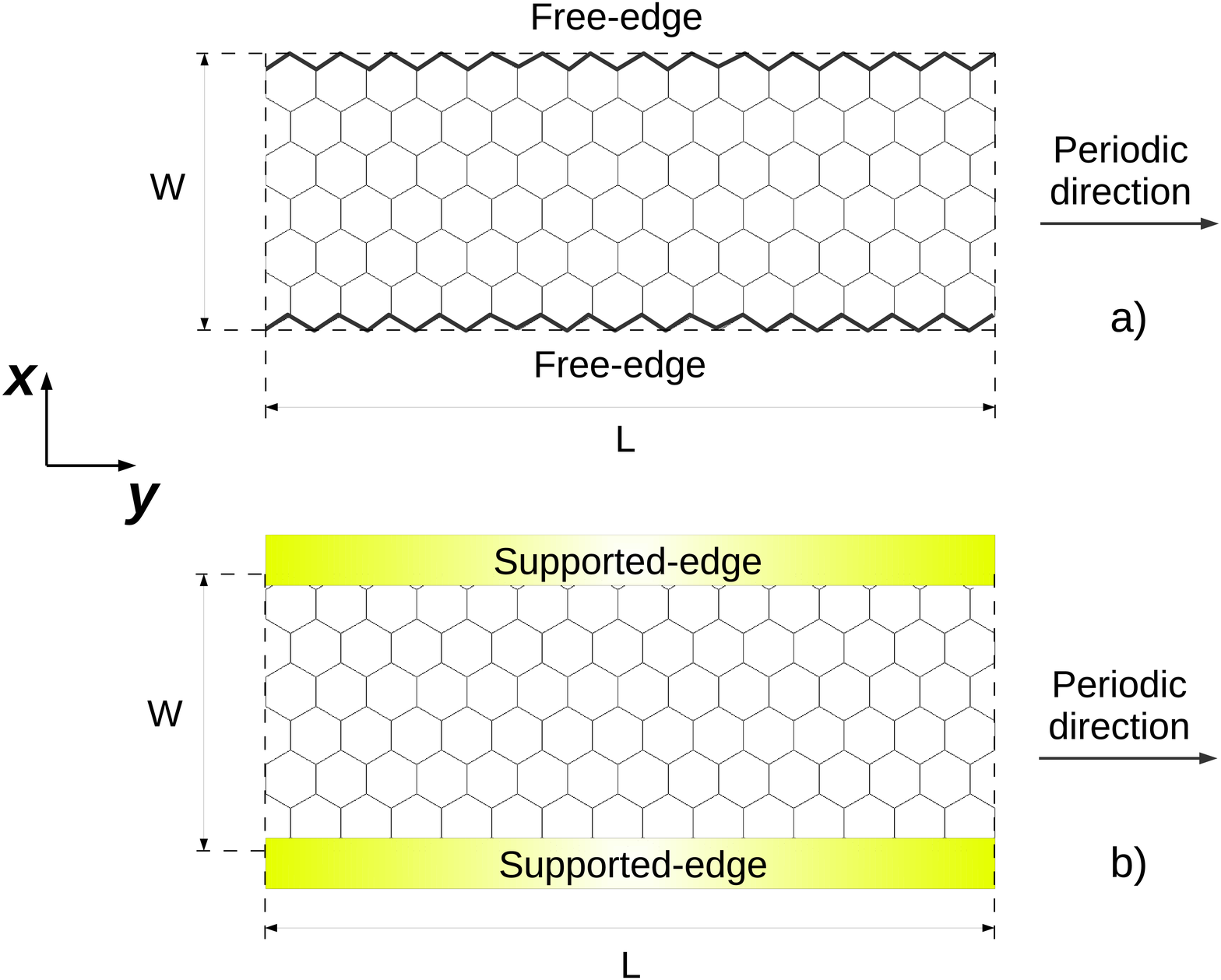}
\caption{ (a) Free-edges and  (b) supported-edges configurations. 
The ribbon width is denoted by W. Periodic boundary conditions are considered in the $y$-direction.}
\label{fig:edges}
\end{figure}

We start by characterizing the phonon modes of GNRs under two 
different boundary conditions: free- and supported-edges (see Fig.~\ref{fig:edges}). As has been proposed in Ref.~[\onlinecite{grapheneacoustic}], both conditions can be realized experimentally. The free-edges configuration corresponds to a strip of graphene that stretches on a wide trench and is supported on the extremes, whereas the supported-edges situation represents a strip that stretches over a thin trench and is supported on the edges.


\subsection{Free-edges GNRs}
\label{sec:free}

{\it Force-constant model.} We consider the harmonic dynamics of carbon (C) atoms in GNRs by using an inter-atomic force-constant model. We employ to this end the 5NNFCM developed 
by M. Mohr and collaborators~\cite{Mohr} and used later by 
Michel and Verberck\cite{Michel} to study the phonon spectrum of single and multi-layer graphene. 
Following the notation adopted in Ref.~[\onlinecite{Michel}], the dynamical matrix can be expressed as:
\begin{eqnarray}
D_{ij}^{\kappa \kappa'}({\bf q})= \sum_{{\bf n}^{\prime}} \frac{\Phi_{ij}({\bf n}\kappa;{\bf n}^{\prime}\kappa^{\prime})}{M} e^{[i {\bf q}\cdot\left({\bf X}({\bf n}^{\prime},\kappa^{\prime})-{\bf X}({\bf n},\kappa)\right)]}
\label{eqn:DM5NNFCM}
\end{eqnarray}
where $M$ is the individual mass of the C atoms, $\kappa=1,2$ reads for the two atoms in each cell (graphene case), 
$i,j=x,y,z$ are the three axis-directions of motion and ${\bf X}({\bf n},\kappa)$ is the 
C atom 2D coordinate in the honeycomb lattice structure.

\begin{figure}[t]
\vspace{0.6cm}
\includegraphics[width=0.48\textwidth]{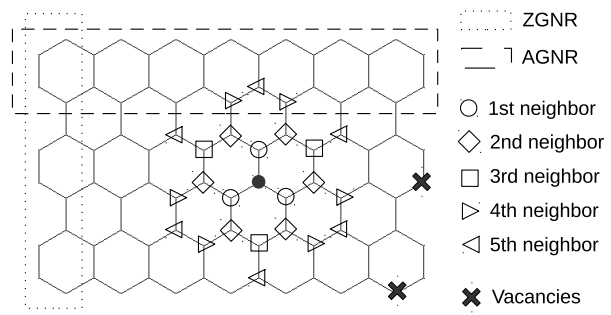}
\caption{Schematic plot of the 5th nearest-neighbor force-constant model 5NNFCM range. 
Dotted (dotted-dashed) lines indicate the unit cell of ZGNRs (AGNRs). Crosses symbolize the vacancies we use when calculating the thermal condunctance in Section III.}
\label{fig2}
\end{figure}
While considering ribbons with a finite number 
of atoms in the transverse direction and adopting periodic boundary conditions in the 
longitudinal direction, the wave-vector $\mathbf{q}$ becomes one-dimensional and 
runs only in the $y$-direction (see Fig.~\ref{fig:edges}).
The force-constant matrix $\Phi_{ij}({\bf n}\kappa;{\bf n}^{\prime}\kappa^{\prime})$ is 
written in terms of parameters which represent the covalent bond forces within the graphene plane\cite{Michel, Mohr}.
A schematic plot of the scope of this potential is displayed in Fig.~\ref{fig2}. 
\begin{figure*}[t]
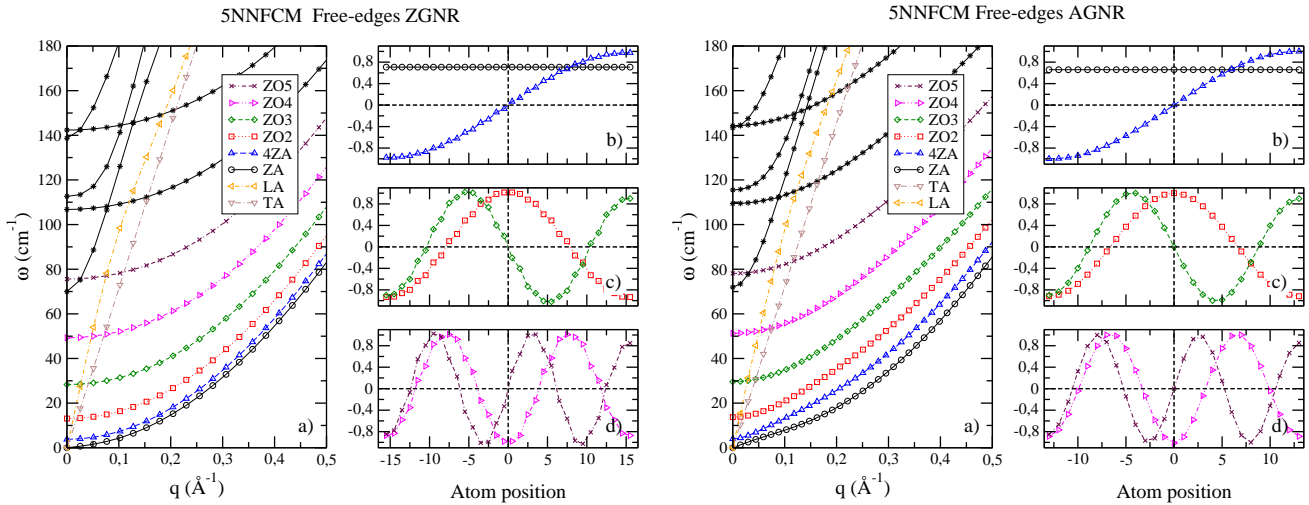

\vspace{0.6cm}
\includegraphics[trim = 0cm 0cm 0cm 0.05cm, clip, width=20pc]{Rel_disp_autov_ZZL32.eps}
\hspace{0.2cm}
\includegraphics[trim = 0cm 0cm 0cm 0.05cm, clip, width=20pc]{RelDisp_autovec_ACL27.eps}
\caption{\label{fig:reldispfree5nnfc} 
a) Dispersion relations for free-edges zigzag (left) and armchair (right) nanoribbons of 32 atoms width calculated with the 5NNFCM. In b), c) and d) we show the corresponding first six eigenmodes.} 
\end{figure*}
The phonon dispersion and the corresponding eigenvectors for a ribbon of finite width W are obtained by diagonalization of the dynamical matrix
$D_{ij}^{\kappa \kappa'}({\bf q})$ given in Eq.~(\ref{eqn:DM5NNFCM}). 
A unit cell consisting in a line (a pair of lines) of C atoms in the transverse direction is adopted 
in order to obtain zig-zag (armchair) edges. 
In Fig.~\ref{fig:reldispfree5nnfc} we show the phonon dispersion and eigenvectors of a free-edges ZGNR (left) and an AGNR (right). In both cases we found the usual three acoustic phonon modes, the ZA (flexural) mode, the TA (transversal) mode and the LA (longitudinal) mode. 
Unlike graphene, GNRs have a rotational symmetry around a central axis in the long direction 
and hence an extra acoustic mode is expected. 
In the present case, the fourth acoustic mode, symbolized as 4ZA in Fig.~\ref{fig:reldispfree5nnfc} (a),
present a small  gap at $\mathbf{q}=0$ (Note that atomic displacements of the 4ZA 
resemble an homogeneous rotation of the ribbon).
We attribute the small gap to the inaccuracy of the 5NNFCM which was defined by fitting graphene bulk phonons 
where this rotational symmetry does not exist. Similar observations for GNRs and CNTs with force-field models and first principles DFT calculations have been reported previously in Refs.~[\onlinecite{Huang, Saito, Yamada, Guillen}].

For AGNRs we find that the properties of the phonon modes are in general very similar 
to the ones obtained for ZGNRs. This is, the main vibrational 
properties at low energies are somehow independent of the microscopic details of the edges.
Note however that very close to $\mathbf{q}$=0, the flexural acoustic ZA mode experiences a change in the 
dispersion from the quadratic behavior to a linear dependence.
A similar trend was obtained by using a force-constant model with interactions up to 4th nearest-neighbors by
Huang {\it et al.} in Ref.~[\onlinecite{Huang}]. 
This behavior is also a consequence of using a potential best suited for graphene.\\  

{\it Elastic theory.}  The vibrational behavior of graphene involving low energy acoustic phonon modes can be studied by the elasticity theory of continuum membranes\cite{Nelson}. The two-dimensional honeycomb lattice is then approximated by a continuum membrane where deformations from the flat configuration are parametrized using the Monge representation $\mathbf{r}(x,y)=(x,y,h(x,y))$, with $h$ being the vertical displacement. The effective free energy $F$ in the harmonic approximation can be expressed as a sum of bending and in-plane elastic energies\cite{Nelson,Landau}:

\begin{eqnarray}
F&=& \int d^2x \frac{\kappa}{2} [(\nabla^2 h)^2 -2Det(\partial_{\alpha}\partial_{\beta}h)]\nonumber \\
 &+& \int d^2x (2\mu u^2_{\alpha \beta} + \lambda u^2_{\alpha \alpha}),
\label{eqn:Etotal}
\end{eqnarray}
where $u_{\alpha \beta}=\frac{1}{2} (\partial_{\alpha} u_{\beta} + \partial_{\beta} u_{\alpha})$ is the strain tensor, $\kappa$ is the bending rigidity and $\mu$ and $\lambda$ the Lam\'e coefficients. 

\begin{figure}[t]
\vspace{0.6cm}
\includegraphics[trim = 0cm 0cm 0cm 0.05cm, clip, width=20pc]{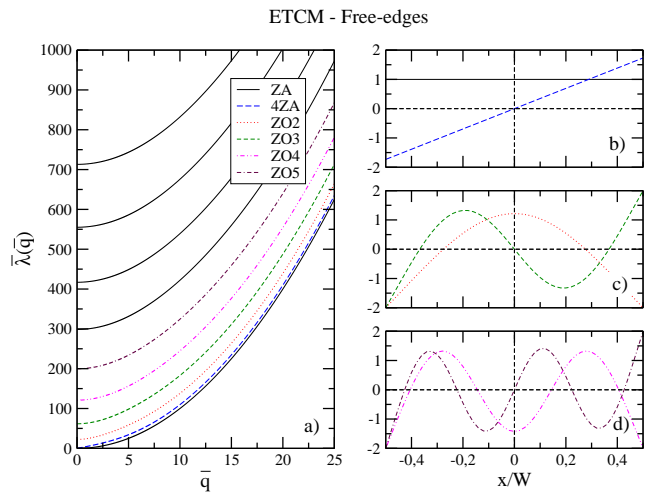}
\caption{ a) Dispersion relation $\bar{\lambda}(\bar{q})=W^2 \sqrt{\frac{\rho}{\kappa}} \omega(qW)$ in terms of $\bar{q}=qW$ for free-edges ribbons. In b), c) and d) we show the eigenfunctions $f_n(x/W)$ for 
the first six branches.} 
\label{fig:reldispmemfree}
\end{figure}
In the harmonic approximation in-plane and out-of-plane modes are completely decoupled. 
In the following we take only into account the out-of-plane deformations. Results concerning in-plane modes were reported in Ref.~[\onlinecite{grapheneacoustic}].
We consider a free-edge ribbon of width $W$ and length $L \gg W$ running along the $y$-direction, 
where we assume periodic boundary conditions.  The edges in the narrow direction, 
located at $x=-W/2$ and $x=W/2$, are taken as free boundaries. 
Then, the coefficients of the variation of $F$ with respect to $h$ and to its 
normal derivative $\partial h /\partial\bf{n}$ at these edges should be zero, which imposes\cite{Landau}: $\left(\partial_x^3 +2 \partial_y^2 \partial_x \right)h(x=\pm\frac{W}{2},y)=0$ and $\partial_x^2 h(x=\pm\frac{W}{2},y)=0$.
Together with the kinetic contribution $\int d^2x \frac{\rho}{2} \dot{h}^2$, 
where $\rho$ is the surface mass density, plane waves of the form $h(x,y,t) =f(x) \, e^{i(q y - \omega t)}$ are proposed as solutions. More details on the calculation can be found in our previous Ref.~[\onlinecite{Scuracchio}].
In Fig.~\ref{fig:reldispmemfree} we show the dispersion relation $\omega_n(q)$ and eigen-functions $f_n(x)$, 
where $n$ indicates the corresponding $n$-mode.
In agreement with results found above with the 5NNFCM, we observe that the lowest energy branch, $n=0$, 
has a quadratic dispersion relation and can be identified with the acoustic flexural mode (ZA). 
Note in Fig.~\ref{fig:reldispmemfree} (b) that $f_0$ is $x$-constant  
and can be understood as an homogeneous translational mode in the $z$-direction. 
The second branch, $n=1$, appears also as a flexural acoustic mode although it is not present in an infinite membrane. 
As we mentioned before, this mode is connected with an additional global rotational symmetry of GNRs. 
The following branches are optical-like and can be thought as overtones of the principal 
modes as we see in Fig.~\ref{fig:reldispmemfree}. 
An overall good agreement is observed with results presented in Fig.~\ref{fig:reldispfree5nnfc}. \\


\begin{figure*}[t]
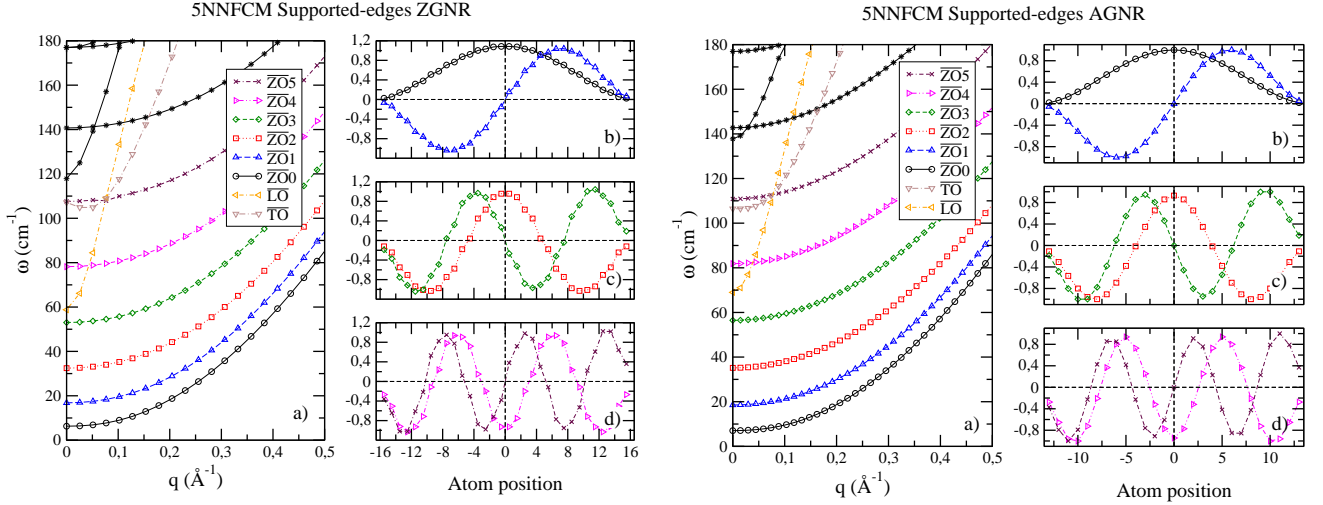

\vspace{0.6cm}
\includegraphics[trim = 0cm 0cm 0cm 0.05cm, clip, width=20pc]{Rel_disp_autov_ZZF32.eps}
\hspace{0.2cm}
\includegraphics[trim = 0cm 0cm 0cm 0.05cm, clip, width=20pc]{RelDisp_autovec_ACF27.eps}
\caption{\label{fig:reldispfix5nnfc} a) Dispersion relations for supported-edges zigzag (left) and armchair (right) nanoribbons of 32 atoms width as given by the 5NNFCM. b,c and d) First six flexural eigenmodes. Note the similarity with the results of a {\bf fixed} elastic membrane.}
\end{figure*}

\subsection{Supported edges}
\label{sec:clampled}

{\it Force-constant model.} The condition of supported-edges  
was set adding fixed extra carbon atoms at both edges extending the honeycomb structure in the $x$-direction. 
Although these extra atoms do not appear as new elements in the lattice Hamiltonian because 
they remain fixed in their equilibrium positions, their interactions modify the diagonal terms according to the 5NNFCM. %
In Fig.~\ref{fig:reldispfix5nnfc} we show the phonon dispersion given by the 5NNFCM for a ZGNR (left) and a AGNR (right).
It can be observed that all the phonon branches have a gap at $\mathbf{q}=0$.
These gaps are related to the energy cost of global rigid displacements since now the edges of the ribbon are fixed. 
Note that the behavior of in-plane modes, labeled as $\bar{LO}$ and $\bar{TO}$, in Fig.~\ref{fig:reldispfix5nnfc} (a) (left and right)
is comparable with results obtained in Ref.~[\onlinecite{grapheneacoustic}] for fixed membranes. 
Out-of-plane modes are further discussed below. 
As a consequence of the supported-edges, the edge atoms are fixed for all eigenvectors and set to zero.
Here, analogously to the case of free-edges ribbons, we observe that overall 
similar trends are obtained between ZGNRs and AGNRs. 

\begin{figure}[t]
\vspace{0.6cm}
\includegraphics[trim = 0cm 0cm 0cm 0.05cm, clip, width=20pc]{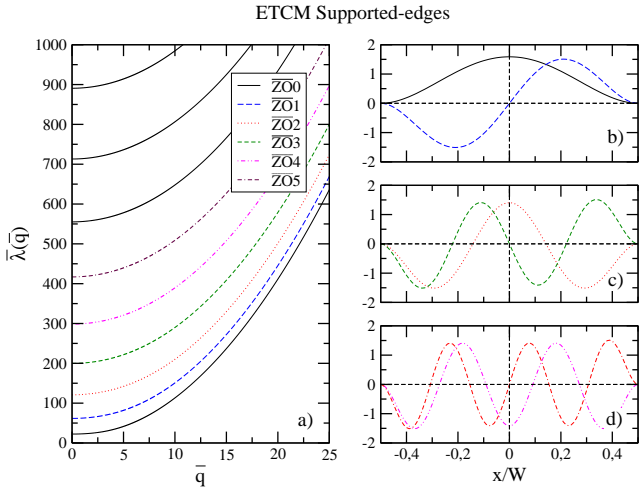}
\caption{a) Dispersion relation $\bar{\lambda}(\bar{q})=W^2 \sqrt{\frac{\rho}{\kappa}} \omega(qW)$ in terms of $\bar{q}=qW$ for supported-edges ribbons b, c y d) Eigenfunctions $f_n(x/W)$ for the first six branches. Note that all eigenfunctions go to zero with no slope at the borders. As free-edges ribbons, the branch number $n$ matches the number of nodes.}
\label{fig:reldispfixmem}
\end{figure}

{\it Elastic theory.} Supported edges in the ETCM are consistent with setting $h(x=\pm\frac{W}{2},y)=0$ and $\partial_xh(x=\pm\frac{W}{2},y)=0$. Physically, one can interpret this 
as having the ribbon sticked to the edge\cite{Scuracchio}. In Fig.~\ref{fig:reldispfixmem} we show the 
phonon dispersion given by the elasticity model under these conditions. A good agreement is found with the results of the 5NNFCM. The main differences with the case of free-edges ribbons is that here a gap appears for all phonon branches 
due to the breakdown of the translational symmetry. 
In Ref.~[\onlinecite{Scuracchio}] we estimated that the gap for the first out-of-plane mode of a supported-edges elastic membrane with $W=30nm$, is around $\sim 7.9 \mu eV$. On the other hand, the gap for the in-plane modes\cite{grapheneacoustic} is around $\sim 1 meV$, being thus highly larger than the former one. As we will show below, this has a strong impact in the thermal transport properties at low temperatures.

\subsection{System size effects}

\begin{figure*}[t]
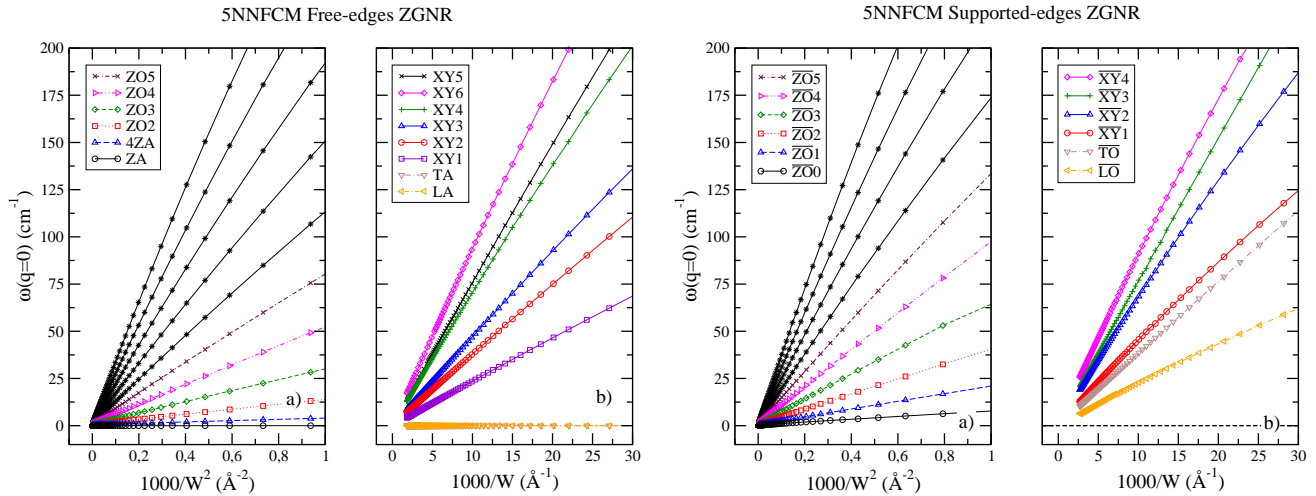

\vspace{0.6cm}
\includegraphics[trim = 0cm 0cm 0cm 0.05cm, clip, width=20pc]{ZZL-Gap-Out-In-plane.eps}
\hspace{0.2cm}
\includegraphics[trim = 0cm 0cm 0cm 0.05cm, clip, width=20pc]{ZZF-Gap-Out-In-Plane.eps}
\caption{\label{fix:gapdependfree} a) Gaps ($\omega(\mathbf{q}=0)$) of out-of-plane modes as functions of width for free-edges ZGNRs (left) and supported-edges ZGNRs (right). b) Same as a) for in-plane modes.}
\end{figure*}

Before discussing thermal transport properties of graphene nanoribbons it is worth to analyze the dependence of 
the energy gap at $\mathbf{q}$=0 with the ribbon width W. In free-edges ribbons, it is naturally expected that when the ribbon width increases, the phonon band-structure of graphene is recovered. Following this reasoning, the larger the ribbon width, the smaller the band-gap of the low-lying optical phonon modes will be.
We have seen that energy dispersion relations and eigenfunctions-eigenmodes present similar behaviors in both, 
continuum elasticity theory and discrete lattice methods. ETCM
predicts~\cite{Scuracchio, grapheneacoustic, Munoz} a scaling of the gaps for in-plane modes 
as $\sim 1/$W and for out-of-plane modes as $\sim 1/$W$^2$. 
In Fig.~\ref{fix:gapdependfree} we show the gap dependence with the ribbon width W 
obtained by the 5NNFCM in ZGNRs. It is observed that both, free- and supported-edges ribbons, 
satisfy the scaling predicted by the elasticity theory. This is important since it allows to 
estimate the band-gaps in large experimental samples. 
Related works found a weakening on the width-dependence 
for in-plane modes with increasing vibrational order in very small samples~\cite{Yamada,Guillen}.
These deviations may arise from hydrogenated edges and geometry relaxation, which have not been included in our study. 

\section{Thermal transport properties}

\subsection{Calculation method}
\label{thermform}

Thermal transport is analyzed in the ballistic regime using the Landauer formalism within the 
NEGF technique~\cite{Zhang3, WangAgarwalla}. Calculations are performed 
in the harmonic approximation neglecting any kind of phonon-phonon or electron-phonon interactions\cite{Mingo,Ghosh}. 
For coherent phonon transport, the atomic structure of graphene nanoribbons can be generalized as 
a central device between two homogeneous semi-infinite contacts where Hamiltonians are defined by using the 5NNFCM
described in Sect.~II. Left and right contacts are thermal reservoirs at constant temperatures $T_1$ and $T_2$ respectively connected to the central device. Following the steps described in Ref.~[\onlinecite{Huang}], surface Green's functions (SGFs) are calculated to obtain phonon vibrational modes at the contact's surfaces. 
These functions are evaluated using the decimation technique\cite{Zhang1,Lopez}
and can be formally expressed as:
\begin{equation}
     g_{L(R)}(\omega)= \lim_{\delta \rightarrow 0} [(\omega^2 +i\delta)I - H_{L(R)C}]^{-1}
  \label{eqn:SGF}
\end{equation}
where  $H_{L(R)CB}$ are the harmonic Hamiltonians of left (right) contacts. 
The introduction of the $\delta$ parameter is directly related to the retarded characteristic of SGFs 
and the computational times involved in the decimation technique~\cite{Mingo,Datta}. 
The convergence criterion adopted here is that the iterative process is finished when 
the largest update of any element of $g_{L(R)}(\omega)$ is less than 1\%. 
Once SGFs are computed, the phonon transmission function is calculated as\cite{Huang,Zhang3}:
\begin{equation}
  \Xi (\omega)= Tr[\Gamma_L(\omega)G(\omega)\Gamma_R(\omega)G^\dagger(\omega)]
  \label{eqn:transmfunc}
\end{equation}
Here $G(\omega)$ is the Green's function of the central device and $\Gamma_{L(R)}(\omega)$ 
is the left (right) broadening matrix. 
The energy flow through the system is evaluated using Landauer formalism:
\begin{equation}
  J=\int \frac{\hbar \omega}{2 \pi} \Xi (\omega) [N(T_L,\omega)-N(T_R,\omega)]  d\omega
  \label{eqn:heatflux}
\end{equation}
where $N(T,\omega)$ is the Bose-Einstein equilibrium distribution of phonons with energy $\hbar \omega$ at temperature $T$. 
If we consider a small temperature difference $T_R-T_L$ between right and left contacts, 
thermal conductance in terms of temperature is defined as
\begin{equation}
  \lambda(T) = \int \frac{\hbar \omega}{2 \pi} \Xi (\omega) \frac{dN}{dT}  d \omega.
  \label{eqn:thermalconduc}
\end{equation}
The transmission function $\Xi (\omega)$ of Eq.~(\ref{eqn:transmfunc}) takes into account the behavior 
of all phonon modes in both contacts. In other words, it can be interpreted as a summation over 
all phonon polarizations at a given frequency $\omega$. In order to calculate different polarization-specific transmissions it is needed to transform the broadening matrices.
Note that this is important since it will allow to determine the effect of atomic vacancies 
and boundary conditions on the transmission of every single phonon mode. 
This procedure has been developed recently in Ref.~[\onlinecite{PolSpecific}]. 
Basically, one has to substitute the broadening matrices $\Gamma_{L(R)}(\omega)$ by 
\begin{equation}
  \gamma_{L(R)} (w)=\sum_i \tau_{L(R)}\lambda_{L(R),i}\varphi_{L(R),i}\varphi_{L(R),i}^\dagger {\tau_{L(R)}^\dagger}
\label{eqn:specificpolgamma}
\end{equation}
where $\{ \lambda_{L(R),i}, \varphi_{L(R),i} \}$ are the corresponding eigenvalues and 
eigenvectors of the matrix $A_{L(R)}=i(g_{L(R)}-g_{L(R)}^{\dagger})$ and $\tau_{L(R)}$ 
are the interaction matrices between left (right) contact and the device. 
By replacing $\gamma_{L(R)} (w)$ for $\Gamma_{L(R)}(\omega)$ in expression (\ref{eqn:transmfunc}), 
we obtain for each value of $i$ the transmission function of 
one polarization mode propagating from left to right (right to left).


\subsection{Boundary conditions}

\begin{figure}[t]
\vspace{0.6cm}
\includegraphics[trim = 0cm 0cm 0cm 0.05cm, clip, width=20pc]{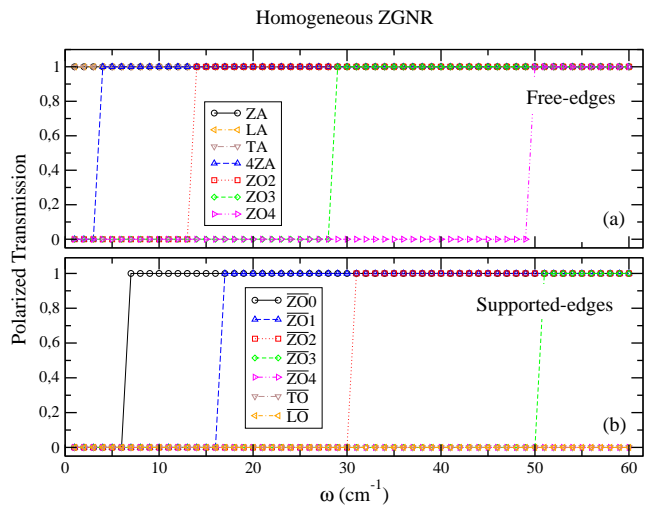}
\caption{Polarized phonon transmission function for (a) free-edges and 
(b) supported-edges ZGNRs. W$=32.6$ \r{A}.}
\label{fig:PurePTM}
\end{figure}

The polarized phonon transmission functions (PPTFs) for the 
free- and supported-edges ZGNRs are shown in Fig.~\ref{fig:PurePTM}. 
As expected for an homogeneous ribbon, the PPTF of each mode is either 0 or 1.
For the free-edge case, we observe the presence of the three acoustic modes ZA, LA and TA 
until the small energy gap of the 4ZA band is reached at $\sim 3 \, $cm$^{-1}$. 
At this point the transmission function value becomes 4 as noticed in Refs.~[\onlinecite{Munoz,Tomita,Datta,TranspRev}] 
for GNRs and CNTs where four acoustic modes are found at $\omega=0$. 
Then, at higher energies in accordance with the phonon spectrum of Fig.~\ref{fig:reldispfree5nnfc} (a, left), 
the extra contributions to the transmission function of the next three optical modes appear exactly 
at the corresponding energies $\omega_1(0)\sim 12 \, $cm$^{-1}$, $\omega_2(0)\sim 28 \, $cm$^{-1}$ y $\omega_3(0)\sim \, 
49 $cm$^{-1}$. Note that the calculations of phonon dispersion modes and transmission functions are independent one of the other.

The case of supported-edges is shown in the Fig.~\ref{fig:PurePTM} (b).
Due to the breakdown in the traslational symmetry, all the phonon modes have an energy gap 
$\omega(0)>$0 and therefore $\Xi (\omega)=0$ until $\omega \sim 6 \, cm^{-1}$ where the first gap is closed. 
This means that no thermal conductance is possible in this energy interval. 
Later on, as in the previous case, the transmission function continuously increases by steps of one unit each time a new gap is overcome as can be seen by comparing with Fig.~\ref{fig:reldispfix5nnfc} (a, left). Although not shown in Fig.~\ref{fig:PurePTM}, $\Xi (\omega)$ also decreases when $\omega$ exceeds the maximum energy of a particular phonon mode.

We mention that while in free-edge GNRs in-plane modes contribute to the thermal 
conductance already from $\omega=0$, in the case of supported-edge GNRs their contribution 
to the transmission function appears only after the contribution of out-of-plane modes.
This can be observed in Fig.~\ref{fig:reldispfix5nnfc} (a, left) where the $\bar{LO}$ mode appears 
at higher energy than the $\bar{ZO}2$ and $\bar{ZO}3$ modes. 
Now, by considering that energy gaps of in-plane modes decay as $W^{-1}$ 
(slower than out-of-plane modes with $W^{-2}$), we obtain that out-of-plane phonon modes will govern only 
the thermal transport at low energy. 
As we show in the following, contributions from out-of-plane phonon modes can be modified by introducing atomic vacancies. 
The behavior of PPTFs for AGNRs under both boundary conditions considered is qualitatively similar. However, it can be seen
that ZGNRs have larger thermal conductances than AGNRs in the entire temperature range\cite{Huang,Tomita,Zhai}.


\subsection{Atomic vacancies}
\label{sec:vacancies}

\begin{figure}[t]
\vspace{0.6cm}
\includegraphics[trim = 0cm 0cm 0cm 0.05cm, clip, width=20pc]{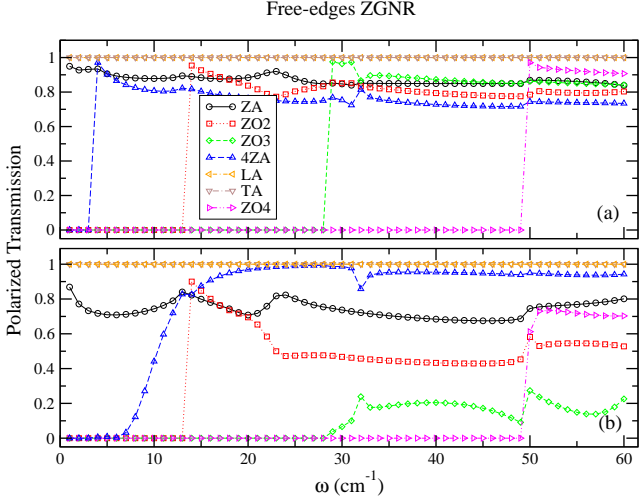}
\caption{Polarized phonon transmission function for a free-edges ZGNR with a single vacancy localized (a) at an edge and (b) at the center.}
\label{fig:TBC-free}
\end{figure}

Let us study the effect of single atomic vacancies located at the edge and at the center of ZZGNRs
under both, free- and supported-edges boundary conditions. Fig.~\ref{fig:TBC-free} shows the PPTF for free-edges ZZGNRs
with a vacancy at (a) the border and (b) the center of the ribbon.
We can observe important differences as compared to the homogeneous case analyzed above.
Here the modes 4ZA and ZO2 deviate considerably from the
perfect transmission value equal to one.
As a rule, all of the out-of-plane contributions become severely diminished by the presence of the vacancy.
In contrast with this, note that the corresponding in-plane contributions remain
absolutely unaffected at low energies. Calculations performed for other widths (not shown here) show similar results.
Following the ideas of Ref.~[\onlinecite{Wang}], this behavior can be related to the respective
group velocity of the given phonon modes. This is, while single atomic vacancies do not interfere
with long-wavelength acoustic phonons with large group velocities, irrespective of their symmetrical properties,
they interact strongly with low speed propagating modes (small group velocity) as the out-of-plane modes.

\begin{figure}[t]
\vspace{0.6cm}
\includegraphics[trim = 0cm 0cm 0cm 0.05cm, clip, width=20pc]{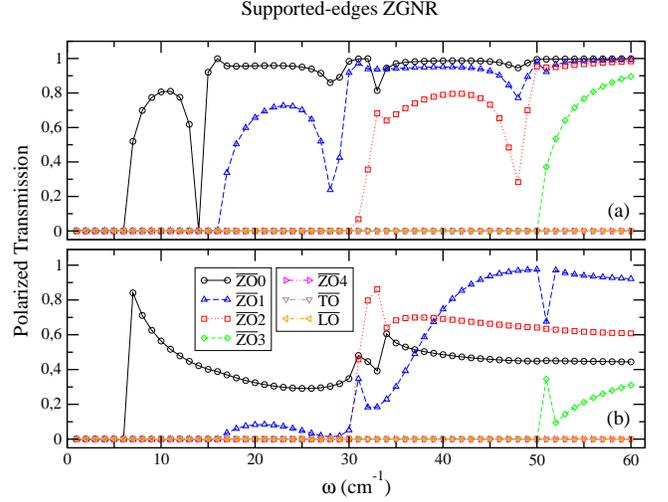}
\caption{Polarized phonon transmission function for a supported-edges ZGNR with a single vacancy localized (a) at an edge and (b) at the center.}
\label{fig:TBC-supp}
\end{figure}

The case of supported-edges ribbons with single atomic vacancies is displayed in Fig.~\ref{fig:TBC-supp}.
The overall behavior for a central vacancy is similar to the case of free-edges, here however
the reductions are somehow larger. 
Note that when the vacancy lies at the center of the ribbon,
irrespective to the edge boundary condition, the contribution of every flexural mode
becomes more affected than when it lies at the edge.
Calculations for other widths show similar qualitative behaviors.


\subsection{Thermal conductance}

\begin{figure}[t!]
\vspace{0.6cm}
\includegraphics[trim = 0cm 0cm 0cm 0.05cm, clip, width=20pc]{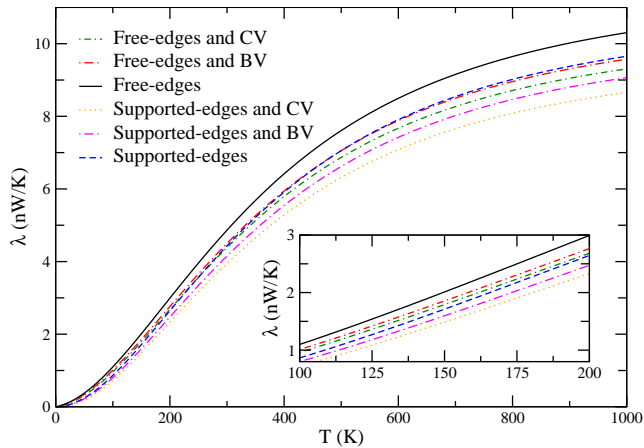}
\caption{Thermal conductance as a function of temperature for different boundary conditions and vacancy localizations. BV stands for a vacancy at the border and CV for a vacancy at the center. In the inset we show a zoom of the thermal conductance for all configurations at low temperatures. }
\label{fig:TConduc}
\end{figure}

Finally, the temperature dependence of thermal conductance for all the above analyzed cases is 
shown in Fig.~\ref{fig:TConduc}. 
In the limit when temperature $T\rightarrow 0$, $\lambda(T)$ approaches the value $3K_0$, where $K_0=\pi^2k_B^2T/3h=(9.456\times 10^{-13}$WK$^{-2}$)T is the thermal conductance quantum\cite{Huang}. The usual result $4K_0$ for 1D systems\cite{TranspRev, Munoz, Tomita, Bae}, such as GNRs and CNTs, is achieved with exactly four acoustic phonon bands. The 4ZA presents a small gap as already discussed, so low temperature behavior does not include this contribution.
Note however that the thermal conductance per unit cross section $\lambda$(T)/(sW) at T=100~K 
calculated with $s=3.5\times10^{-10}$~m and W$=3.26$~nm (free-edge ZGNR of Fig.~\ref{fig:TConduc})
is consistent to those reported in Refs.~[\onlinecite{Munoz, Bae, Pop2}] 
where $\lambda(T)/(sW)\sim 10^9$ WK$^{-1}$m$^{-2}$.

According to the PPTFs of previous sections, we observe that the largest $\lambda$(T), in the whole range of temperatures, 
is obtained for free-edges GNRs.
The introduction of vacancies reduces $\lambda$(T) due to the lower contributions of flexural phonons.
As we discussed above, by supporting the ribbon at its edges, a large gap opens up for in-plane phonons, 
which produces a further reduction on $\lambda$(T). 

An interesting crossover takes place at low temperatures (inset in Fig.~\ref{fig:TConduc}). 
Because of the large gap of in-plane phonon modes 
in supported-edge ribbons, $\lambda$(T) is smaller 
as compared to the free-edges ribbons with vacancies either at the border or at the center.
At higher temperatures, the situation is reverted and in-plane phonon modes 
start to contribute to $\lambda$(T). This process can be understood by 
noticing that at high temperatures, homogeneous supported-edges 
ribbons conduct heat via both, in-plane and out of plane phonons modes, while free-edges ribbons 
with vacancies have reduced (not perfect) contributions from flexural phonons.
Therefore, by introduction of vacancies or supporting the ribbon at its edges 
we have different mechanisms to control the relative contributions of phonon modes
to the thermal conduction at different temperatures. 

In the ballistic regime, when the width of the ribbon 
increases extra phonon modes are available for the thermal conductance and 
$\lambda$(T) increases. Although not shown here, we found that the mentioned 
crossover holds for diferent ribbons widths.
Beyond the ballistic regime, nonmonotonic behavior of the thermal conductivity $\kappa$ 
in terms of lateral dimensions and edge roughness are expected~\cite{Anomalous}.


\section{Conclusions}

The effects of edge and central located single atomic vacancies on the ballistic 
thermal conductance of graphene nanoribbons was calculated by means of the Landauer formalism\cite{Zhang3} using NEGF methods. 
We analyzed the cases of free- and supported-edges ribbons. 
As a first step, we provided a full comparison of the vibrational phonon modes  
obtained by a fifth-nearest-neighbor force-constant model 5NNFCM, used before in the description of graphene\cite{Michel}, and the continuum elasticity theory. 

We analyzed in detail the microscopical characteristics of in-plane and out-of-plane low energy phonon modes 
and we ascribed the finite energy edge-localized fourth acoustic phonon mode 4ZA, obtained through the 5NNFCM, 
to the lack of rotational symmetry. This effect is also present in DFT-based results\cite{Guillen}.  
We demonstrated that the scalings against the ribbon width W 
of the phonon band-gaps follow $1/$W and $1/$W$^2$ laws for in-plane and out-of-plane modes, respectively.

We found that in-plane polarized transmissions can be strongly reduced at 
low energy by setting fixed boundary conditions. On the other hand, 
single vacancies reduce considerably only the out-of-plane polarized transmissions. 
We believe that these findings could open new routes to design graphene based structures 
where a control of the polarized mode transmission is possible, thus having important consequences in thermal transport of nanoscale-devices.


\section{Acknowledgments}

Discussions with S. D. Dalosto and K. H. Michel are gratefully acknowledged. 
This work was partially supported by PIP 11220090100392 of CONICET (Argentina) and the Flemish Science Foundation (FWO-VI).
We acknowledge funding from the FWO~(Belgium)-MINCyT~(Argentina) collaborative research
project. 


\end{document}